\documentclass[aps,prx,twocolumn,showpacs,superscriptaddress,floatfix,nofootinbib]{revtex4-1}
\usepackage{amsfonts}
\usepackage{graphicx}
\usepackage{amsmath}
\usepackage{times}
\usepackage{amssymb}
\usepackage{txfonts}
\usepackage{color}
\usepackage[colorlinks,bookmarks=false,citecolor=blue,linkcolor=red,urlcolor=blue]{hyperref}

\begin{document}

\title{
Spin-orbital quantum liquid on the honeycomb lattice
}

\author{Philippe Corboz}
\affiliation{Theoretische Physik, ETH Zurich, 8093 Zurich, Switzerland}

\author{Mikl\'os Lajk\'o}
\affiliation{Institute for Solid State Physics and Optics, Wigner Research
Centre for Physics, Hungarian Academy of Sciences, H-1525 Budapest, P.O.B. 49, Hungary}
\affiliation{Department of Physics, Budapest University of Technology and Economics, 1111 Budapest, Hungary}

\author{Andreas M. L\"auchli}
\affiliation{Institut f\"ur Theoretische Physik, Universit\"at Innsbruck, A-6020 Innsbruck, Austria}

\author{Karlo Penc}
\affiliation{Institute for Solid State Physics and Optics, Wigner Research
Centre for Physics, Hungarian Academy of Sciences, H-1525 Budapest, P.O.B. 49, Hungary}
\affiliation{Department of Physics, Budapest University of Technology and Economics, 1111 Budapest, Hungary}

\author{Fr\'ed\'eric Mila}
\affiliation{Institut de th\'eorie des ph\'enom\`enes physiques, \'Ecole Polytechnique F\'ed\'erale de Lausanne (EPFL), CH-1015 Lausanne, Switzerland}

\date{\today}

\begin{abstract}
In addition to low-energy spin fluctuations, which distinguish them from band insulators, Mott insulators often possess orbital degrees of freedom when crystal-field levels are partially filled. While in most situations spins and orbitals  develop long-range order, the possibility for the ground state to be a quantum liquid opens new perspectives. In this paper, we provide clear evidence that the SU(4) symmetric
Kugel-Khomskii model on the honeycomb lattice is a quantum spin-orbital liquid. The absence
of any form of symmetry breaking - lattice or SU(N) - is supported by a combination of semiclassical
and numerical approaches: flavor-wave theory, tensor network algorithm, and exact diagonalizations.
In addition, all properties revealed by these methods are very accurately accounted for by a projected variational wave-function based on the $\pi$-flux state of fermions on the honeycomb lattice at $1/4$-filling. In that state, correlations are algebraic because of the presence of a Dirac point at the Fermi level, suggesting
that the symmetric Kugel-Khomskii model on the honeycomb lattice is an algebraic quantum spin-orbital liquid.
This model provides a good starting point to understand the recently discovered spin-orbital liquid behavior of Ba$_3$CuSb$_2$O$_9$.  The present results also suggest to choose optical lattices with honeycomb geometry in the search for quantum liquids in ultra-cold four-color fermionic atoms.
\end{abstract}


\maketitle

\section{Introduction}
The investigation of orbital physics in transition metal oxides has been
recently boosted by the possibility to observe orbital excitations with
Resonant Inelastic X-ray Scattering\cite{schlappa2012}. This has been demonstrated in cases
where the crystal-field splitting is strong enough to select a unique orbital
configuration in the ground state and push the orbital excitations to high
energy, hence to separate them from magnetic excitations. However, this is not
the only  possibility. If the electronic configuration of the transition metal
ion is such that several orbital occupations are consistent with the crystal
field environment, a situation referred to as orbital degeneracy\cite{brink2011}, orbital
fluctuations are likely to be much softer. In most cases known until recently, a
cooperative Jahn-Teller distortion occurs, resulting in orbital order and
gapped orbital excitations, but this needs not be the case a priori, and
the search for situations in which orbitals remain fluctuating in the ground
state has been very active over the past decade~\cite{ishihara1997,feiner1997,li1998,khaliullin2000, vernay2004,wang2009,chaloupka2011}. To which extent this occurs in the
triangular system LiNiO$_2$ \cite{kitaoka1998,reynaud2001} or in the spinel FeSc$_2$S$_4$ \cite{fritsch2004} is still debated~\cite{mostovoy2002}.
Interestingly, a new candidate has been recently put
forward, Ba$_3$CuSb$_2$O$_9$~\cite{nakatsuji2012}, a Cu oxide that lives on a decorated honeycomb
lattice in which no trace of orbital order could be detected.

On the theory side, transition-metal oxides with orbital degeneracy are generally
described by a Kugel-Khomskii model~\cite{kugel1982} in which spin and orbital degrees of freedom
are coupled on each bond. A minimal model to investigate the possibility to stabilize
an orbital liquid is the symmetric version of that model defined by the Hamiltonian
\begin{equation}
{\cal H}=\sum_{\langle i,j\rangle} (2 \mathbf{S}_i \cdot \mathbf{S}_j + \frac{1}{2})(2 \mathbf{T}_i\cdot \mathbf{T}_j + \frac{1}{2})
\label{eqn:P2Ham}
\end{equation}
where $\mathbf{S}_i$ is a spin-1/2 operator and $\mathbf{T}_i$ a pseudo-spin 1/2 operator that
describes fluctuations of a two-fold degenerate orbital ($a$ and $b$).
Introducing the local basis $\vert {\color{red}\medbullet}  \rangle = \mid \uparrow a \rangle$,
$\vert {\color{green}\medbullet} \rangle = \mid \downarrow a \rangle$, $\vert {\color{blue}\medbullet} \rangle = \mid \uparrow b \rangle$,
$\vert {\color{yellow}\medbullet} \rangle = \mid \downarrow b \rangle$, the Hamiltonian exhibits the full SU(4) symmetry and can be written as $\mathcal H = \sum_{\langle i,j\rangle}  P_{i,j}$, where  $ P_{i,j}$ interchanges the states on sites $i$ and $j$.
The local basis states are often referred to as colors. In fact, the model is the straightforward generalization of the SU(2) symmetric Heisenberg model for $S=1/2$ spins which, up to a constant and a factor 2, has the same form when expressed with $ P_{i,j}$ operators.


\begin{figure*}[]
\includegraphics[height=6.5cm]{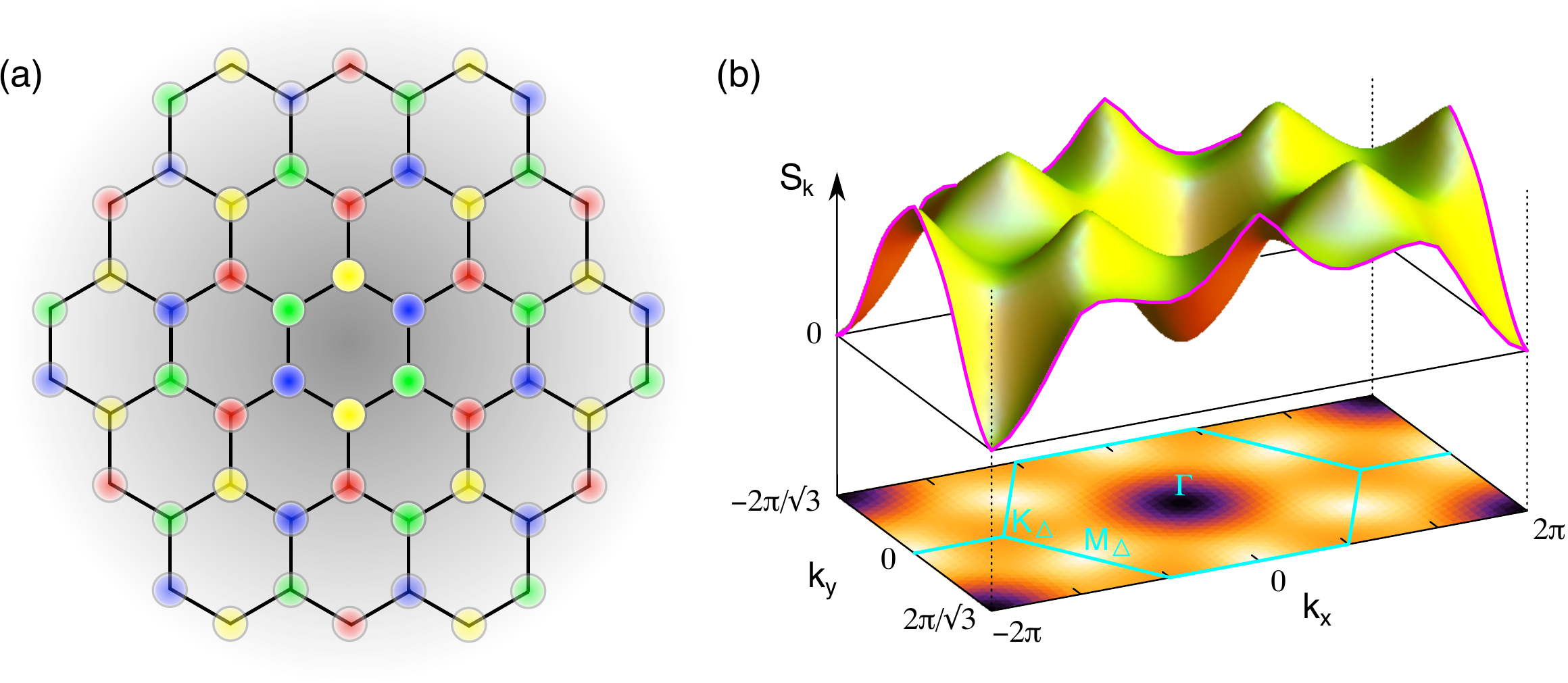}
\caption{Summary of the main properties of the spin-orbital liquid of the SU(4) model on the honeycomb lattice.
(a) Sketch of the local color-order. This pattern is the only one that respects the
sequence of 4 colors along all the zigzag chains, whatever their orientation (horizontal, $\pi/3$
or $2\pi/3$). (b) Color structure factor of the Gutzwiller projected $\pi$-flux state (VMC). The singular conical peaks are typical of algebraic correlations.}
\label{fig:states}
\end{figure*}

The first investigations of this model on various lattices have emphasized the role of 4-site
plaquettes, the natural unit to build an SU(4) singlet~\cite{li1998, van_den_bossche2000, van_den_bossche2001}. The spontaneous formation of 4-site
plaquettes has been proven for an SU(4) ladder~\cite{van_den_bossche2001}, and plaquette coverings have been argued to provide
the relevant variational subspace for the ground state properties of the SU(4) model on both
the square and triangular lattices~\cite{li1998,van_den_bossche2000}, with possibly plaquette long-range order on the square lattice~\cite{hung2011,szirmai2011}. In a variational study based on projected fermionic wave functions a gapless spin-orbital liquid state has been predicted in Ref.~\cite{wang2009} on the square lattice.
These previous conclusions have recently been challenged  for
the square lattice~\cite{corboz2011}, for which spontaneous dimerization (as opposed to tetramerization) has been
demonstrated on the basis of state-of-the-art iPEPS (infinite projected entangled-pair state)
simulations. 
In the dimerized phase, the dimers are antisymmetric states built out of 2 of the 4 colors, and they form a
columnar state. Each dimer is preferentially
surrounded by dimers built out of the other 2 colors, leading to long-range color order \cite{corboz2011}.
In the context of orbital degeneracy, this phase
is likely to be ordered. Indeed, one of the states selected by the dimerization consists
of alternating pairs of $a$ and $b$ orbitals times a spin singlet. When coupled to the lattice,
such a state is expected to undergo a cooperative Jahn-Teller distortion that will
stabilize orbitals $a$ and $b$, hence to lead to orbital long-range order.

In this paper, we consider the symmetric Kugel-Khomskii model on the honeycomb lattice.
The first motivation is purely theoretical: since there are no 4-site plaquettes on this
lattice, the ground state is unlikely to be a crystal of singlet plaquettes. The second one
comes from experiments: the recent observation of a spin-liquid behaviour in Ba$_3$CuSb$_2$O$_9$
points to the honeycomb geometry as an outstanding candidate. 
%
%

The main result of the present
investigation is summarized in Fig.~\ref{fig:states}: The SU(4) symmetric Kugel-Khomskii model
is shown to be a quantum spin-orbital liquid with short-range color correlations that follow
the pattern illustrated in
the top panel, and strong evidence is provided in favor of an algebraic spin-orbital liquid with
typical conical singularities in the static structure factor, as shown in the bottom panel.

\begin{figure}[]
\begin{center}
\includegraphics[width=8.5cm]{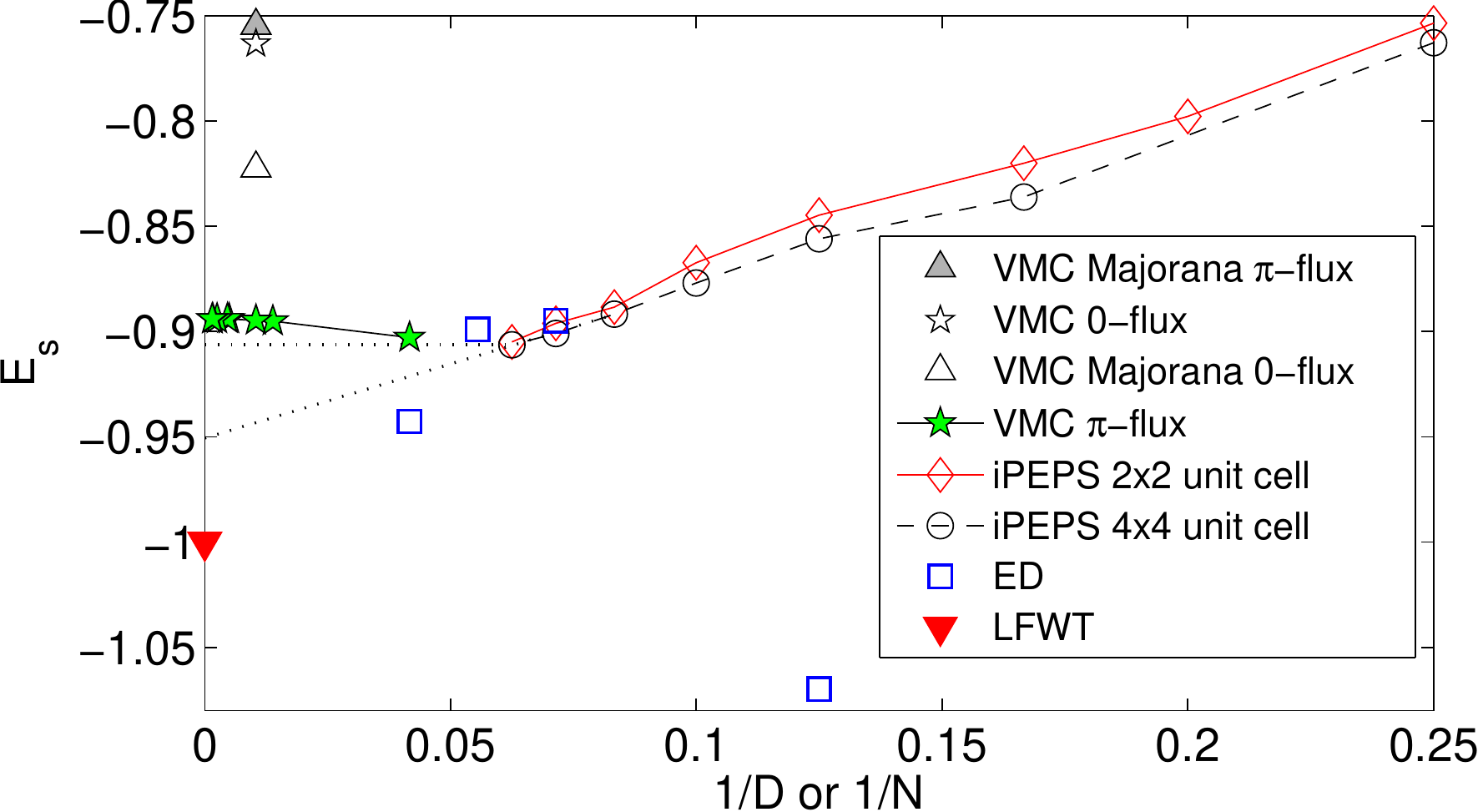}
\caption{Energy per site as a function of inverse bond dimension $D$ (iPEPS) and as a function of inverse system size $N$ (VMC and ED). We note that the LFWT energy is \emph{not} variational.
}
\label{fig:res_iPEPS}
\end{center}
\end{figure}

To reach these conclusions, we have used a variety of analytical and numerical methods: Linear flavor-wave theory (LFWT),
infinite projected entangled-pair states (iPEPS), exact diagonalization of finite clusters (ED), and a variational approach based on the Monte Carlo sampling of Gutzwiller projected fermionic wave-functions (VMC). Details about each method can be
found in appendix~\ref{sec:methods}. These methods are complementary and shed lights on different aspects of the model:
LFWT is a good starting point to test for lattice symmetry breaking and color order.  iPEPS is a variational
approach for infinite systems that has proven to be very successful to check the presence of any kind
of long-range order~\cite{corboz2011, corboz2012}. Exact diagonalizations reveal nearly exact information on short length-scale properties and are extremely useful to benchmark other approaches. The variational Monte Carlo of fermionic
wave-functions have proven to provide a remarkably accurate description of algebraic quantum liquids.
As a first test, we compare in Fig.~\ref{fig:res_iPEPS} the ground state
energies of the various approaches. A number of conclusions can already be drawn from this comparison. First
of all, among all the Gutzwiller projected fermionic wave-functions we considered, only one is really competitive, the wave-function
based on the quarter-filled Fermi sea of standard fermions in the $\pi$-flux state (see Sec.~\ref{sec:asol} for details). Its energy is much lower
than that of the 0-flux state, as well as that of the half-filled Fermi sea of Majorana fermions with 0 or $\pi$
flux, and these alternative fermionic wave-functions will not be considered any further\footnote{The variational energies (per site) of the $\pi$-flux, Majorana $0$-flux,  $0$-flux, and  Majorana $\pi$-flux states in the 96 site cluster are $-0.895$, $-0.822$, $-0.763$, and $-0.755$, respectively.}.
Secondly, the agreement between iPEPS, ED and Gutzwiller projected $\pi$ flux state is quite remarkable. This suggests that all these
methods constitute appropriate descriptions within their range of validity.\footnote{LFWT leads to a very low energy.
This is not a concern since the method is not variational, but this is an indication that, in the present
case, higher orders are likely to be important.}

\subsection{Absence of lattice symmetry breaking}

\begin{figure}[]
\begin{center}
\includegraphics[width=8cm]{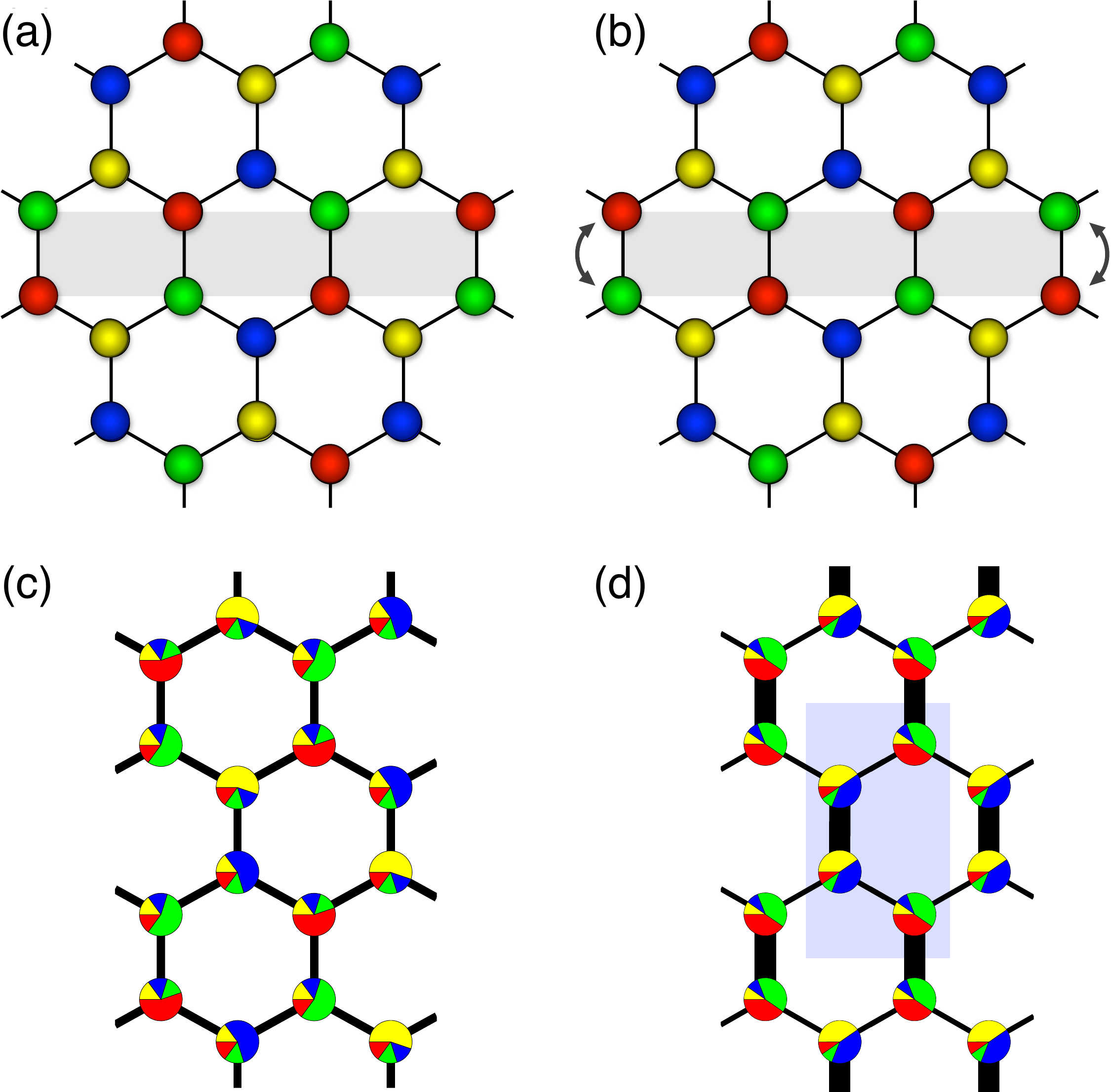}
\caption{
Examples of states obtained with LFWT (a-b) and iPEPS for small bond dimension $D$ (c-d). 
(a) Most symmetric configuration. (b) Configuration obtained from the most symmetric one by exchanging colors on a stripe.
(c) Color-ordered state with one dominant color per site, obtained with a $4\times4$ unit cell and a bond dimension $D=6$.  (d)~Dimer-N\'eel ordered state obtained with a $2\times 2$ unit cell (shaded in blue) and $D=6$. Both the color order and the dimerization vanish in the infinite $D$ limit. 
The pie charts show the local color density on each site and the thickness of a bond is proportional to the square of the energy on the corresponding bond.
}
\label{fig:iPEPS_states}
\end{center}
\end{figure}

Lattice symmetry breaking, be it dimerization or plaquette formation, leads
to bonds of different strengths. At the classical level, which consists in minimizing
the energy in the subspace of product wave-functions of the form
$\vert \psi \rangle = \prod_i \vert \psi_i \rangle$, all bonds are fully satisfied.
Indeed, since the Hamiltonian of a bond $H_{ij}$ is a simple permutation, $\langle \psi_i \psi_j \vert H_{ij} \vert \psi_i \psi_j \rangle = \vert \langle \psi_i \vert \psi_j \rangle \vert^2$ is minimal if neighboring
states are orthogonal. On bipartite lattices such as the square or honeycomb lattices, it takes only
2 colors to achieve this, and the classical ground state for more than 2 colors is massively
degenerate. This degeneracy however can be in principle lifted by zero-point fluctuations. The theory
of harmonic fluctuations has been developed previously, and it goes under the name of flavor-wave theory
(see appendix \ref{sec:lfwt} for details). At the harmonic level, the energy of a bond takes the smallest
possible value if the two colors of the bond are different from the other nearest neighbors.
For the honeycomb lattice, this can be achieved for all bonds simultaneously in an infinite
number of ways. The configuration of Fig.~\ref{fig:states}(a) is the most symmetric one. In this configuration, all bonds have the same surrounding
up to color permutations. Other configurations can be generated by exchanging the colors on a
stripe of dimers (see Fig.~\ref{fig:iPEPS_states}(a-b)), leading to a degeneracy of order $2^{\sqrt{N}}$ since, once a
direction has been chosen, this can be done independently on all dimer stripes. In all configurations,
the energy is the same on all bonds. This is a first hint that,
by contrast to the square lattice, the lattice symmetry is not broken for the honeycomb lattice.
%

\begin{figure}[]
\begin{center}
\includegraphics[width=8.5cm]{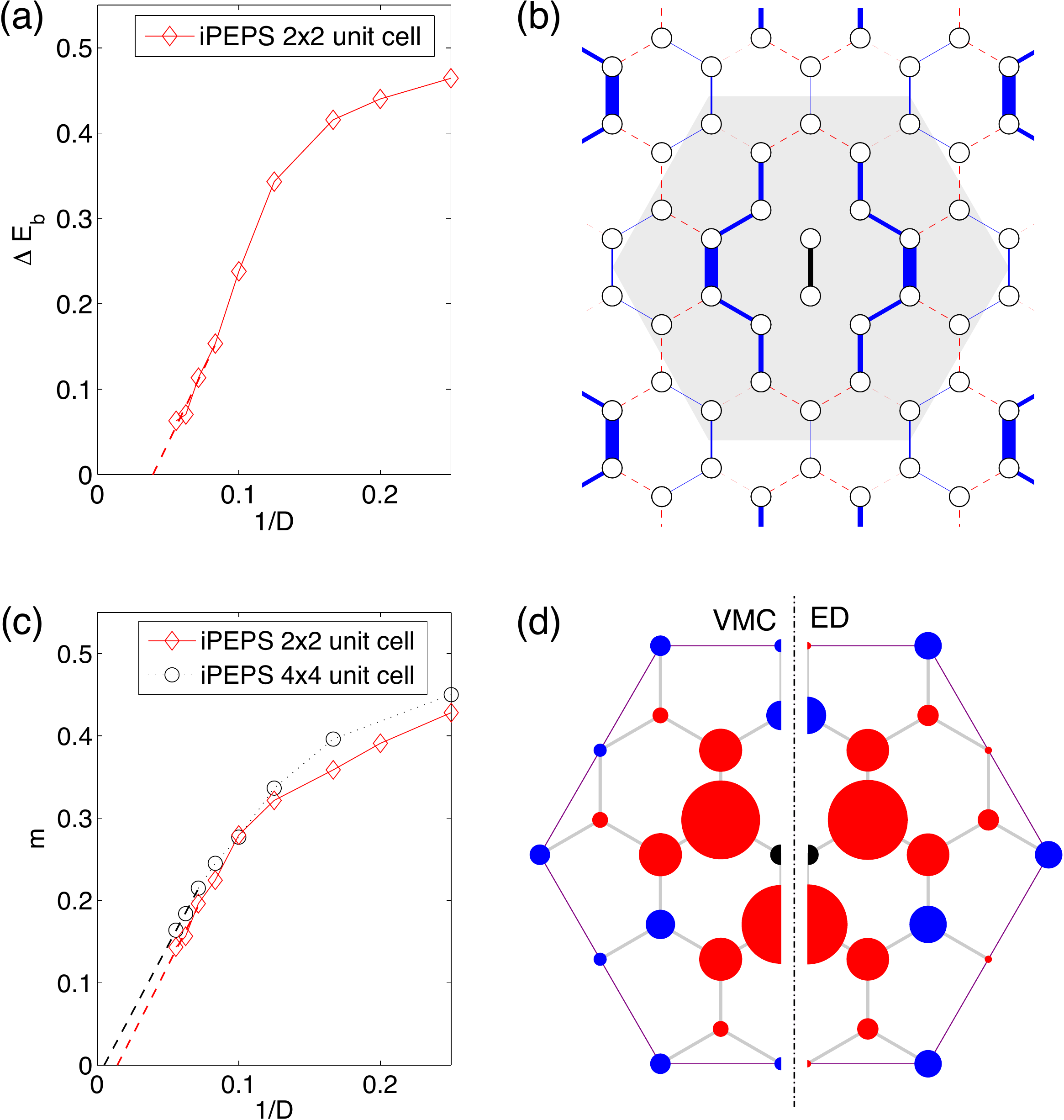}
\caption{
Various correlation functions obtained with iPEPS, ED and VMC. 
(a) Difference in bond energies obtained with iPEPS in the state shown in Fig.~\ref{fig:iPEPS_states}(d). It is strongly suppressed with increasing bond dimension $D$ and vanishes in the infinite $D$ limit.
(b) Connected bond energy correlations $\langle P_{ij}P_{kl}\rangle-\langle P_{ij}\rangle\langle P_{kl} \rangle$ calculated with
ED in the ground state of the $N=24$ sample. The black bond denotes the reference bond. Solid blue (dashed red) bonds stand
for positive (negative) correlation functions, and the width is proportional to the absolute value of the correlation function.
(c) Local ordered moment $m$ obtained with iPEPS as a function of inverse bond dimension. It vanishes in the infinite $D$ limit for both low-energy states shown in Fig.~\ref{fig:iPEPS_states}(c-d).
 (d)  Spin correlation function in real space, as calculated from ED (right) and VMC (left) for a 24-site cluster. The area is proportional to $\langle P_{0i} \rangle-1/4$, where 0 is the index of the central site, and $i$ labels the sites in the 24-site cluster. The color keeps track of the sign (blue for positive, red for negative).}
\label{fig:sym}
\end{center}
\end{figure}

The same family of degenerate ground states is obtained with iPEPS if a unit cell with $4\times 4=16$ different tensors and a small bond dimension $D=2$ are used. Upon increasing $D$, more quantum fluctuations are taken into account, and the symmetric state of Fig.~\ref{fig:iPEPS_states}(a) is stabilized. In this state, all bonds have the same energy (see Fig.~\ref{fig:iPEPS_states}(c)).
To test how robust this conclusion is, we have challenged it by performing iPEPS simulations using a $2\times 2$ unit cell with only four different tensors. This leads to a dimerized state with two types of dimers A and B, which can be distinguished by their dominant colors, and different inter- and intra-dimer bond energies (Fig.~\ref{fig:iPEPS_states}(d)).
However, unlike on the square lattice,
this dimerization vanishes in the infinite $D$ limit, as shown in Fig.~\ref{fig:sym}(a) where the difference in bond energies, $\Delta E_b = \max(E_b) - \min(E_b)$, is plotted. Thus, both low-energy states found with iPEPS preserve the lattice symmetry in the large $D$ limit.

We also tested with VMC the stability of the $\pi$-flux state towards the dimerization instability shown in Fig.~\ref{fig:iPEPS_states}(d) by strengthening/weakening the hopping amplitude of the bonds, with the conclusion that the variational energy is minimal in the absence of any dimerization.
Similarly, we found no indication toward quadrumerization (SU(4) singlet formation), where the hoppings connected to say red sites (forming a `tripod') in Fig.~\ref{fig:states}(a) are modified.

Finally, in Fig.~\ref{fig:sym}(b) we show the ED results for the connected bond energy correlations, which is a way to detect dimerization and other bond energy pattern formation tendencies. The correlations decay quite rapidly with distance, making dimerization or other patterns unlikely.

Thus, all methods consistently point towards a state which does not break the lattice symmetry.

%

\subsection{Absence of SU(4) symmetry breaking}
The color-ordered states predicted by LFWT and iPEPS with a small bond dimension (Fig.~\ref{fig:iPEPS_states}) break the SU(4) symmetry. Here we show that higher-order quantum fluctuations destroy this color order,  i.e. that in the ground state the SU(4) symmetry is in fact unbroken.

In Fig.~\ref{fig:sym}(c) we present the iPEPS result for the local ordered moment,
\begin{equation}
\label{eq:m}
 m=\sqrt{\frac{4}{3} \sum_{\alpha,\beta} \left(\langle S_\alpha^\beta \rangle - \frac{ \delta_{\alpha\beta}}{4 } \right)^2},
\end{equation}
where $S_\alpha^\beta = |\alpha\rangle\langle \beta|$ are the generators of SU(4) and $\alpha,\beta$ run over the four different flavors. A finite $m$ implies that the SU(4) symmetry is spontaneously broken in the thermodynamic limit. For both low-energy states found with iPEPS  the local ordered moment is strongly suppressed with increasing bond dimension, and most likely vanishes in the large $D$ limit. This shows that quantum fluctuations, which are systematically taken into account by increasing $D$, eventually destroy the color order so that the SU(4) symmetry is restored. This is in contrast to the same model on the square lattice \cite{corboz2011} where the local ordered moment was found to remain finite in the infinite $D$ limit.

Consistent results for the flavor correlation function are obtained with ED and VMC for the $\pi-$flux state  shown in Fig.~\ref{fig:sym}(d), which is decaying rapidly with increasing distance, indicating absence of long range order. The very good qualitative and quantitative agreement between the ED and the VMC results provides substantial evidence that the $\pi-$flux state correctly describes the short range physics of the ground state of the Hamiltonian \eqref{eqn:P2Ham}. 
In the next section we will show that the decay predicted by VMC is algebraic, i.e. that the
state described by this wave function is an algebraic spin-orbital liquid.

\begin{figure}[]
\begin{center}
\includegraphics[width=8.5cm]{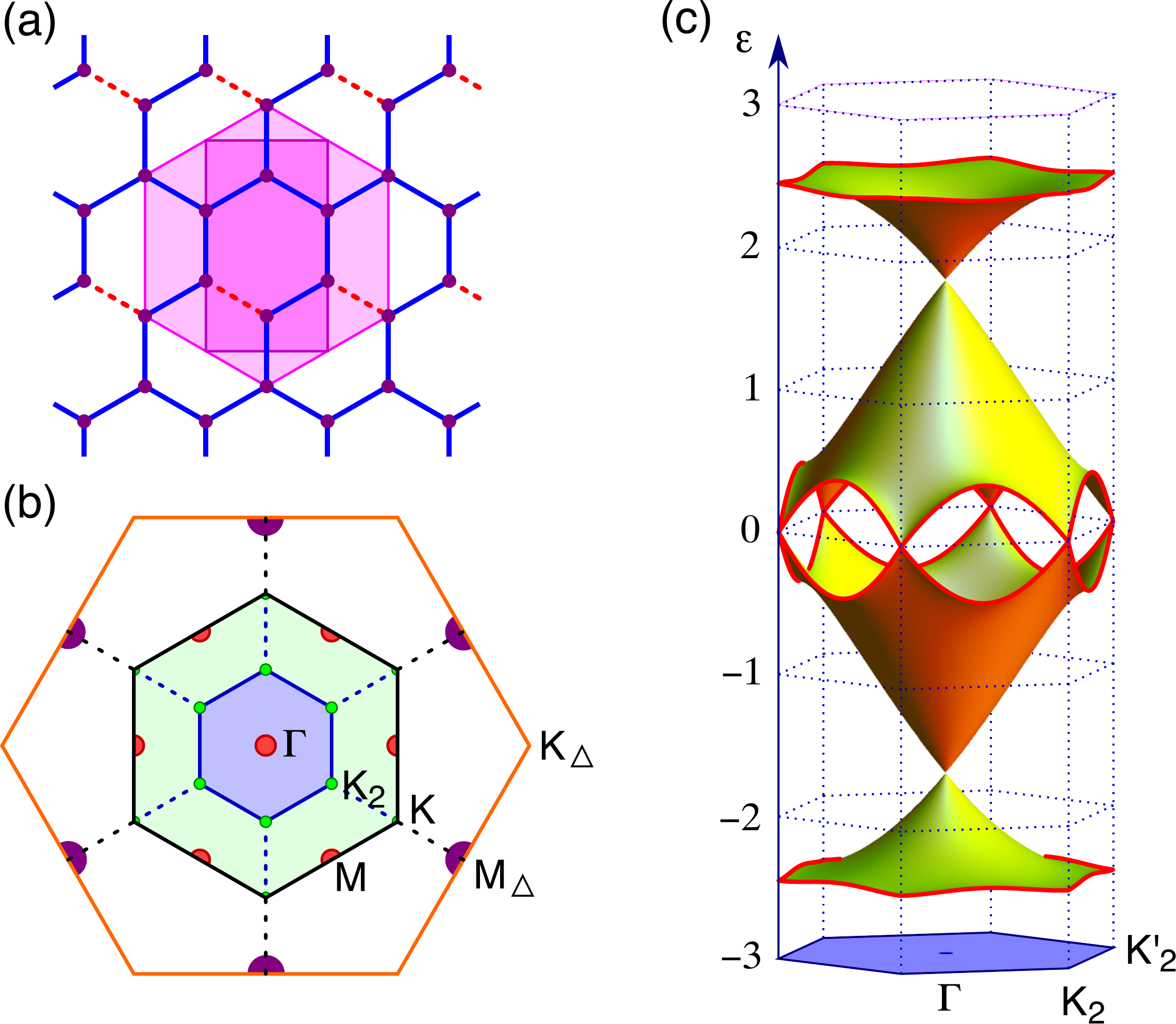}
\caption{Properties of the $\pi$-flux state state. 
(a) Sketch of the gauge used to implement the $\pi$-flux state: the hopping amplitude is positive on solid
blue bonds, negative on dashed red bonds. The primitive unit cell (dark magenta) contains four sites, the hexagonal unit cell eight sites.
(b) Brillouin zones and high symmetry points. The red circles indicate the position of the Dirac-nodes at $\varepsilon_{D}=\pm \sqrt{3}t$ to which the Fermi surface reduces at quarter filling in the $\pi$-flux state. The orange, outermost hexagon shows the extended Brillouin zone of the triangular lattice (including sites at the centers of the hexagons in the honeycomb lattice), the structure factor is maximal and has a cusp at $M_\triangle=(\pi,\pi/\sqrt{3})$ and the symmetry related points. $K_\triangle$ is given by $(4\pi/3,0)$, $K$ is $(2\pi/3,2\pi/3\sqrt{3})$.
(c) The two-fold degenerate band structure of the $\pi$-flux state in the reduced Brillouin zone of 8-site hexagonal unit cell.
}
\label{fig:bandstructure}
\end{center}
\end{figure}

\subsection{Algebraic spin--orbital liquid}
\label{sec:asol}
A standard way to describe spin liquids for SU(2) models is based on the fermionic 
representation of the spin operators \cite{wen2002,motrunich2005,hermele2005,ran2007,paramekanti2007} using a variational wave function
\cite{yokoyama1987,gros1989}, where the multiply occupied sites are projected out from a suitable chosen, noninteracting Fermi-sea. 
In the generic case, the Fermi-sea has a finite Fermi surface. However this is not the only possibility. For the SU(2) Heisenberg model on the square lattice, Marston and Affleck have shown that introducing a  $\pi$-flux per elementary plaquette leads to the formation of Dirac-nodes \cite{affleck1988}.
At half filling, the Fermi surface of this $\pi$-flux state shrinks to points, and its energy is lower than that of the state with equal hopping amplitudes and a finite Fermi surface. In such a spin-liquid the structure factor is singular at momenta related to the difference between Fermi-points, leading to the algebraic decay of spin correlations. 
In one dimension, this type of approach leads to an accurate description of the algebraic decay of the correlations for the SU(2) case \cite{kaplan1982}, and as well as for SU(4) using the representation  $S_\alpha^\beta = f^{\dagger}_\alpha f^{\phantom{\dagger}}_\beta$~\cite{wang2009}.
%


On the honeycomb lattice, a Dirac-node is already present at the middle of the band without any flux, and the Fermi surface reduces to points at half filling. So the zero flux state would be a good starting point to describe an
algebraic spin liquid for the SU(2) Heisenberg model~\cite{hermele07,albuquerque11,clark11}. However, for the SU(4) Heisenberg model, the band must be
quarter-filled, and the equivalent of the Affleck-Marston approach requires to have a Dirac node at the Fermi energy of the quarter-filled system. It turns out that this is achieved in the $\pi$-flux state, as shown in Fig.~\ref{fig:bandstructure}(c).
As for the square lattice, this state leads to a lower energy than the zero-flux state, as already stated above.

Starting from the noninteracting wave function, with a band populated up to the Dirac-node at $\varepsilon_{D}=-\sqrt{3}t$ for any of the four flavored fermions, we implemented the Gutzwiller-projection using a variational Monte-Carlo (VMC) sampling. The energy of this wave function, $E = -0.894$ per site, compares remarkably well with that of iPEPS (see Fig.~\ref{fig:res_iPEPS}), especially considering that no variational parameter was used. Let us also mention that the state (and the ones related by symmetry) shown in Fig.~\ref{fig:states}(a) has the maximal weight in the variational wave function.

\begin{figure}[]
\begin{center}
\includegraphics[width=8.5cm]{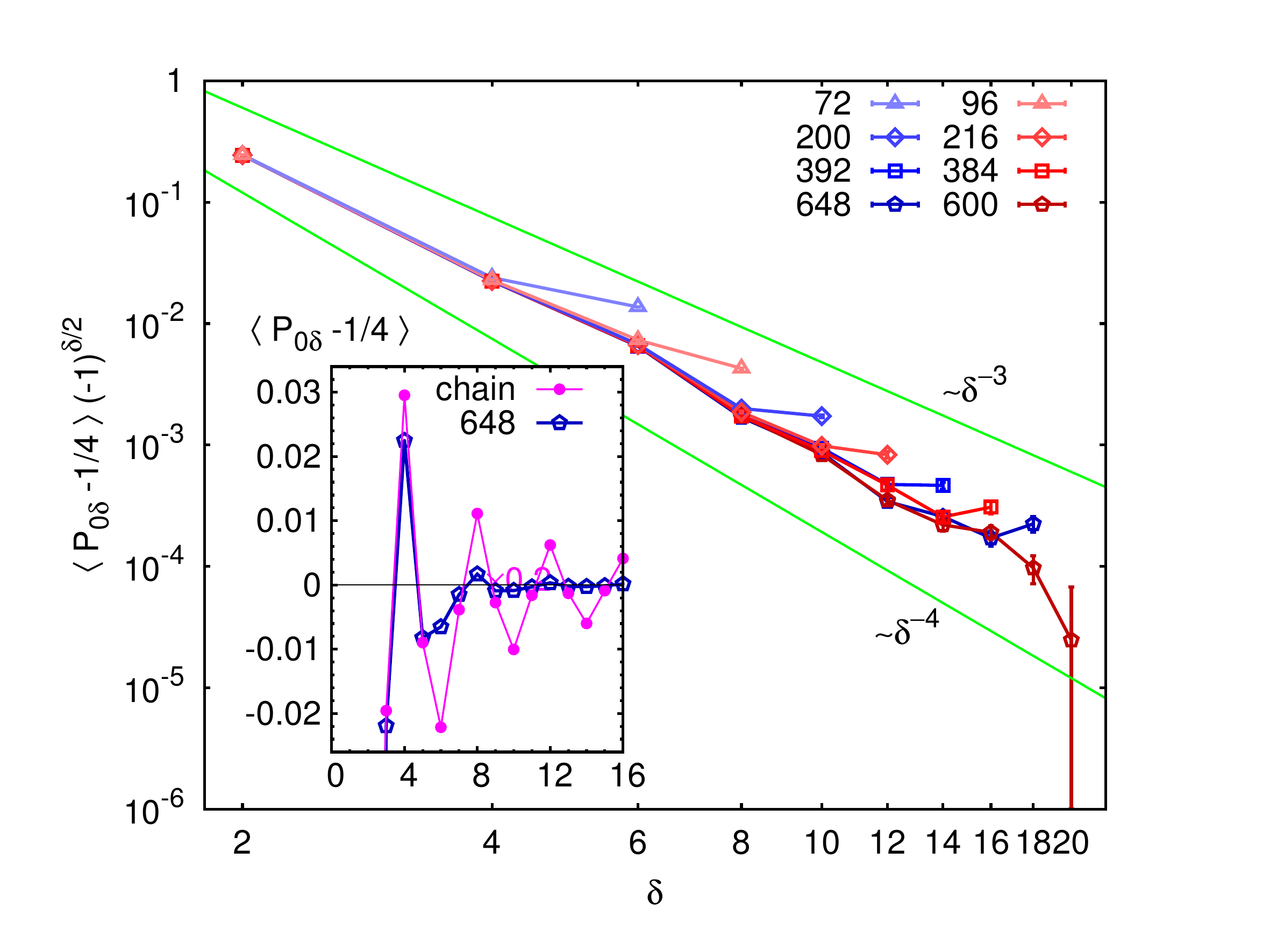}
\caption{The algebraic decay of the correlation along a zig-zag chain in the honeycomb lattice for the Gutzwiller projected $\pi$-flux state in different clusters (every second site is shown, $\delta$ is the distance along the chain). In the inset we compare the correlations of the same $\pi$-flux state with the correlations of a 300 site long one-dimensional chain (we projected the one-dimensional quarter-filled Fermi-sea). While the periodicity of four is visible in both cases, the correlations decay much faster in the two-dimensional honeycomb lattice.
}
\label{fig:algebraic_decay}
\end{center}
\end{figure}

To investigate the physics of this wave function, we have calculated the spin-spin correlation function as a function of distance. The results clearly demonstrate an algebraic decay $\langle P_{ij} - 1/4\rangle \sim |{\bf r}_{ij}|^{-\alpha}$, with an exponent $\alpha$  between 3 and 4, as shown in Fig.~\ref{fig:algebraic_decay}.
If one considers the honeycomb lattice as built from zigzag chains, these correlations correspond to even distances along one of the zigzag chains, and the exponent should be compared to that of the dominant correlations with wave vector $\pi/2$ of a single chain~\cite{Frischmuth1999}.
This exponent is equal to 3/2, a number actually very accurately reproduced by VMC. So color-color correlations decay
faster on the honeycomb lattice than on a chain, but still algebraically. This is a rather peculiar situation in view of the standard paradigms: the development of long-range order, as in weakly coupled SU(2) chains in square geometry, or the spontaneous formation of local singlets and exponentially decaying color-color correlations, as e.g. in the SU(4)  ladder~\cite{van_den_bossche2001}.

This Gutzwiller projected $\pi$ flux state is actually a prototypical wave function for a phase which should be called algebraic spin-orbital liquid in the present context, in analogy to the algebraic spin liquids which have been discussed in the spin liquid literature~\cite{wen2002,hermele2005,assaad2005,ran2007}. These states are characterized by the algebraic decay of a number of correlations, with wave vectors corresponding to differences between the Dirac cone locii. In our case for example the algebraic spin-orbital correlations are modulated with wave vector $M$ in the standard honeycomb Brillouin zone, and this corresponds to the distance between two Dirac cones in the $\pi$ flux state. Another important aspect of this wave function is that it has been shown that it can describe an extended phase in
parameter space and not just an unstable fine-tuned point~\cite{rantner2001,hermele2005}. Upon adding perturbations of suitable strength, many different phases can be found in the vicinity of an 
algebraic spin-orbital liquid~\cite{hermele2005}, making the present model an interesting starting point for further explorations of exotic phases in spin-orbital systems.


\section{Conclusions}

In conclusion, the results reported in this paper provide very strong evidence that the SU(4) symmetric
Kugel-Khomskii model is a quantum spin-orbital liquid, and build a case in favor of an algebraic spin-orbital
liquid. Clearly, the present results do not allow to exclude that the quantum spin-orbital liquid is of
another type, for instance some kind of resonating valence-bond (RVB) liquid with resonances between 4-site
cluster singlets, but our results suggest that the corresponding gap would be quite small. In particular, experience with highly frustrated magnets have shown that, as good
as its energy might be, a fermionic variational wave function might fail to capture the correct low-energy physics.
This seems for instance to be the case for the SU(2) Heisenberg model on the kagome lattice, for which
the variational energy of the algebraic spin liquid wave function~\cite{ran2007} is close to the best numerical estimates~\cite{yan2011,Laeuchli2011}, yet DMRG results have given strong evidence in favor of a gapped $Z_2$ spin liquid~\cite{yan2011}. In the present case, the projected fermionic wave function has not just been
tested for its energy, but correlations have been shown to be in remarkable agreement with ED up to intermediate distances. So we believe that the case is strong, but not closed.

In any case, the fact that the ground state is a quantum spin-orbital liquid is quite firmly established. This
result is quite interesting in view of the liquid behavior reported recently in Ba$_3$CuSb$_2$O$_9$. A detailed
comparison will clearly require some adaptation of the present model, in particular to take into account
the additional magnetic Cu sites present in the system on top of the honeycomb lattice. Still, the absence
of orbital order reported in the present paper is consistent with the experimental results.

Finally, we note that the SU(N) Heisenberg model is a rather accurate effective model for the $1/N$-filled Mott insulating phase of alkaline-earth metal atoms with N internal degrees of freedom loaded in an optical lattice~\cite{cazalilla2009,hermele2009, gorshkov2010}.
Currently the main issue in that field is to reach low enough entropies to observe correlations typical of long-range order,
but the next step will definitely be to realize exotic quantum states. In that respect, the $N=4$ case on the
honeycomb lattice appears to be a very strong candidate.

\section{Acknowledgments}

The ED simulations have been performed on machines of the platform "Scientific computing" at the University of
Innsbruck - supported by the BMWF, and the iPEPS simulations on the Brutus cluster at ETH Zurich.
We thank the support of the Swiss National Science Foundation, MaNEP, and the Hungarian OTKA Grant No. K73455.

\begin{appendix}

\section{Methods}
\label{sec:methods}
\subsection{Linear flavor-wave theory}
\label{sec:lfwt}
The linear flavor-wave theory (LFWT) is a method to treat harmonic quantum fluctuations on top of a mean-field (or Hartree) solution based on a site-factorized variational wave function~\cite{papanicolaou1984,joshi1999}. It starts from a Schwinger boson representation of the SU(4) operators with 4 types of bosons. In a mean-field ground state, each site has a well defined color. The corresponding boson is assumed to condense, and the resulting Hamiltonian is a bosonic quadratic form. On a nearest-neighbor bond with color $\alpha$ on site $i$ and color $\beta$ on site $j$, it is given by
$\mathcal{H}_\text{fw}= A^{\dagger}_{ij} A^{\phantom{\dagger}}_{ij} -1$, with $A^{\dagger}_{ij} = b^{\dagger}_{\alpha,j} + b^{\phantom{\dagger}}_{\beta,i}$
where $b^\dagger$ and $b$ are bosonic operators.  In a given mean-field ground state, the LFWT Hamiltonian is the sum of independent Hamiltonians that describe the motion of bosons on the connected clusters spanned by pairs of colors. The zero-point energy per bond tends to increase with the cluster size, and is minimal on a 2-site cluster, when the ground state energy of the Hamiltonian is equal to $-1$, i.e., there is no zero-point contribution to the energy. By contrast, larger clusters lead to finite frequencies, hence to strictly positive contributions to the zero-point energy.

\subsection{Infinite projected entangled-pair states (iPEPS)}
A projected entangled-pair state (PEPS)~\cite{verstraete2004,verstraete2008}, also called tensor product state, is a variational ansatz where the wave function of a two-dimensional system is efficiently represented by a product of  tensors, with one tensor per lattice site. It can be seen as a two-dimensional generalization of matrix product states - the class of variational states underlying the famous density matrix renormalization group (DMRG) method~\cite{white1992}. On the square lattice each tensor $T^p_{ijkl}$ has a physical index $p$ carrying the local Hilbert space of a lattice site with dimension $d$, and four auxiliary bond indices $i, j, k, l$ with dimension $D$ which connect to the four neighboring tensors. Thus, each tensor consists of $dD^4$ variational parameters, and by changing the bond dimension $D$ the accuracy of the ansatz can be systematically controlled.  
A $D=1$ PEPS simply corresponds to a site factorized wave function (a product state), and upon increasing $D$ quantum fluctuations can be systematically taken into account.

\begin{figure}[]
\begin{center}
\includegraphics[width=8.5cm]{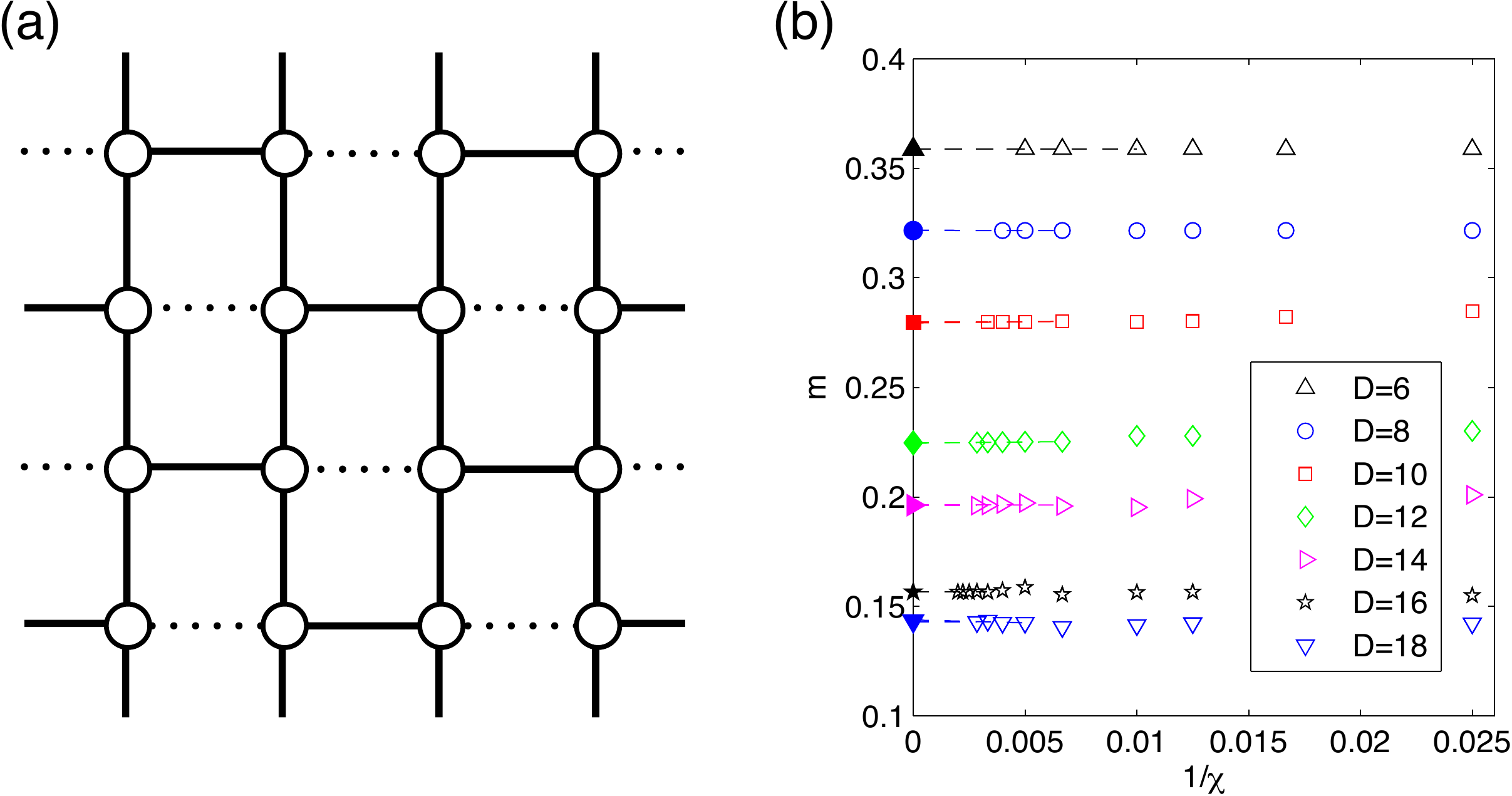} 
\caption{(a) The honeycomb lattice is mapped to a square lattice with brick-wall structure. There is no Hamiltonian term between sites connected by a dotted line. A square lattice iPEPS is used for this lattice, where we choose the bond dimension along the dotted lines as $D=1$. (b) Ordered moment as a function of inverse $\chi$ which controls the accuracy of the contraction of the iPEPS. The values of $m$ for different bond dimensions $D$ only slightly depend on $\chi$.
}
\label{fig:brickwall}
\end{center}
\end{figure}

An infinite PEPS (iPEPS) \cite{jordan2008,corboz2011-1} consists of a unit cell of $L_x \times L_y=N_T$ tensors which is periodically repeated in the lattice to represent a wave function in the thermodynamic limit. We use the iPEPS method developed for the square lattice described in Refs.~\cite{jordan2008, corboz2010, corboz2011} to simulate the model on the honeycomb lattice by mapping it onto a  brick-wall lattice as illustrated in Fig.~\ref{fig:brickwall}(a). The bond dimension of the auxiliary bonds connecting two sites which do not directly interact (dotted lines) can be chosen as $D=1$. This ansatz is equivalent to an iPEPS with only three auxiliary indices on the honeycomb lattice. The advantage of this mapping is that the codes developed for the square lattice only require minor modifications to simulate models on the honeycomb lattice. 

Describing the iPEPS method in full detail is beyond the scope of this paper and we therefore only mention  the most important technicalities with corresponding references and details on the simulation parameters in the following. For an introduction to PEPS and iPEPS we refer to Refs.~\cite{verstraete2008,corboz2010}.

The optimization of the tensors (i.e. finding the best variational parameters) is done through imaginary time evolution, first with the so-called simple update (see Refs.~\cite{jiang2008,corboz2010}) which is equivalent to  the time-evolving block decimation method in one dimension \cite{vidal2003-1, orus2008}. The solution is used as an initial state for an imaginary time evolution using the full update \cite{jordan2008, corboz2010}, which is computationally more expensive, but more accurate than the simple update, since the full wave function is taken into account for the truncation of a bond index. We use a second order Trotter-Suzuki decomposition with time steps down to $d\tau=10^{-3}$. For large values of $D$ a larger time step $d\tau=10^{-2}$ is used, where the estimated Trotter error is small compared to symbol sizes (e.g. below $0.5\%$ for the ordered moment $m$ \eqref{eq:m}).

Expectation values of observables can be computed by contracting the tensor network, i.e. by computing the trace of the product of all tensors. For the approximate contraction of the iPEPS we use the corner transfer matrix method described in Refs.~\cite{nishino1996, orus2009-1,corboz2010}. The accuracy of this contraction can be controlled by the so-called boundary dimension $\chi$, where we used values up to $\chi=500$ (typically up to $\chi=350$) for large $D$.  Observables like the energy and the ordered moment are extrapolated in $\chi$, with an extrapolation error being small compared to symbol sizes. An example is shown in  Fig.~\ref{fig:brickwall}(b) for the ordered moment  $m$ \eqref{eq:m}.

To increase the efficiency of the method we implemented a $\mathbb{Z}_q$ symmetry in the tensors (see Refs.~\cite{singh2010-1,bauer2011}) which is a discrete subgroup of SU(4). This implies that the tensors have a block structure (similar to a block diagonal matrix) which reduces the computational cost considerably.

Since iPEPS is an ansatz for an infinite system, symmetries such as SU(4) or translational symmetry may be spontaneously broken. In order to test different types of translational symmetry breaking we compared the variational energies obtained with different unit cell sizes. We found two competing low-energy states with unit cell sizes $2\times2$ and $4\times4$, shown in Fig.~\ref{fig:iPEPS_states}, which have similar energies for large bond dimension. 
We note that broken symmetries can be an artifact of a finite bond dimension $D$, and that the symmetry can be restored if $D$ is sufficiently large. In other words, a classical or a low entanglement solution (small $D$) might exhibit order, but this order can be destroyed by quantum fluctuations which are systematically included upon increasing $D$. Therefore, it is important to study order parameters as a function of $D$ to verify if they are finite in the large $D$ limit. 

\subsection{Exact diagonalization}

We have performed exact diagonalizations of the Hamiltonian \eqref{eqn:P2Ham} for selected finite size samples of up to $N=24$ sites.
The energies of the samples with 8,14,18 and 24 sites are reported with star symbols in Fig.~\ref{fig:res_iPEPS}. Note that only the
samples with 8 and 24 sites can form a SU(4) singlet ground state, and this explains the relatively high energy per site for the samples with
14 and 18 sites. Note also that due to computational limitations we were only able 
to calculate the energy and eigenfunction of the ground state in the
completely symmetric representation of the spatial symmetry group of the $N=24$ sample (the Hilbert space in this symmetry sector contains 4'008'417'658 states, 
including a cyclic color permutation symmetry).
Since this sector is the absolute ground state for the $N=8$ sample, we expect this sector  to host the ground state for $N=24$ as well.

\subsection{Fermionic Variational Monte Carlo}

The variational wave function for the algebraic spin-orbital liquid is
a Gutzwiller projected noninteracting Fermi sea at quarter filling:
\begin{equation}
|\Psi\rangle = \sum_{\{j\}}
\prod_{\alpha=1}^{4} w_{\{j^{\alpha}\}} |j_1^{\alpha}j_2^{\alpha}\dots j_{N/4}^{\alpha} \rangle
\end{equation}
where $j_l^{\alpha}$ denotes the position of the $l$-th fermion with color $\alpha$, and the summation is over the all  $ N!/ ((N/4)!)^4$ possible distributions of the fermions so that each site is single occupied (i.e. $\{j^{\alpha}\}\cap\{j^{\beta}\}$ is an empty set for $\alpha \neq \beta$). The weight of each configuration is given by the
\begin{equation}
w_{\{j^{\alpha}\}} =
\left|
\begin{array}{cccc}
\xi_1(j_1^{\alpha})  & \xi_2(j_1^{\alpha})  & \dots  & \xi_{N/4}(j_1^{\alpha}) \\
 \vdots &  \vdots &\ddots & \vdots  \\
\xi_1(j_{N/4}^{\alpha})  & \xi_2(j_{N/4}^{\alpha})  & \dots  & \xi_{N/4}(j_{N/4}^{\alpha}) \\
\end{array}
\right|
\label{eq:wjdef}
\end{equation}
Slater-determinant, where $\xi_k(j)$ is the amplitude of the fermion at site $j$ in the $k$-th eigenfunction of the hopping Hamiltonian
\begin{equation}
 \mathcal{H}_{f}= - \sum_{\langle i,j \rangle} \sum_{\alpha=1}^4 \left(\chi^{\phantom{\dagger}}_{i,j} f_{j,\alpha}^{\dagger}f_{i,\alpha}^{\phantom{\dagger}} +
 \text{h.c.}
 \right) \;.
  \label{eq:Hmf}
\end{equation}
The expectation values of operators with this variational wave function are evaluated by a Monte Carlo sampling \cite{yokoyama1987,gros1989}. 
 The variational wave function with Majorana fermions is described in detail in Ref.~\cite{wang2009}.

 The different trial states correspond to different choices of the $\chi_{i,j}$ hopping parameters. In the $\pi$-flux state $|\chi_{i,j}|=1$ and the phases are chosen so that an electron hopping around the hexagon picks up a minus sign, $\prod_{k=1}^6 \chi_{i_{k},i_{k+1}} =-1$, where the product is over the bonds of a hexagon. This can be realized in many ways. Here we choose real hopping amplitudes, where every hexagon has a single bond with a $\chi_{ij}=-1$, while the rest of the bonds have  $+1$, as shown in Fig.~\ref{fig:bandstructure}(a). Furthermore, we allow for antiperiodic boundary conditions when a degeneracy of quarter filled Fermi sea needs to be removed for a given cluster.

We considered two families of finite size clusters with the full $D_6$ symmetry of the honeycomb lattice:
(i) clusters with $N = 2(2n)^2$ sites defined by the lattice vectors ${\bf g}_1 = (\sqrt{3},0)n$ and ${\bf g}_2 = (\sqrt{3}/2,3/2)n$, where $n$ is an integer ($N=72$, 200, 392, and 648 site clusters);
(ii) clusters with $N = 6 (2n)^2$ sites  with ${\bf g}_1 = (3\sqrt{3}/2,3/2)n$ and ${\bf g}_2 = (0,3)n$, ($N=$24, 96, 216, 384, 600).

 We use Monte Carlo sampling of the projected wave function $|\Psi\rangle$ to evaluate physical quantities. 
The elementary step is the exchange of two randomly chosen fermions with different colors. To speed up the convergence we used importance sampling, with acceptance ratios defined according to the Metropolis algorithm: the new configuration is always accepted if its weight $w_{\text{new}}$ is higher than the weight $w_{\text{old}}$ of the old configuration, but for $w_{\text{new}} < w_{\text{old}}$ it is accepted only with a probability $w_{\text{new}}/w_{\text{old}}$.
The configurations are thus represented with a probability $p_{\{j\}} $ proportional to their weight in the wave function:
\begin{equation}
  p_{\{j\}} \propto |w_{\left\{j\right\}}|^2,
\end{equation}
where $w_{\left\{j\right\}}= \prod_{\alpha=1}^{4} w_{\{j^{\alpha}\}} $ denotes the coefficient of $ | \{j\}\rangle =\otimes_{\alpha=1}^4 |j_1^{\alpha}j_2^{\alpha}\dots j_{N/4}^{\alpha} \rangle$ in the projected fermionic  state $ | \Psi \rangle$, see Eq.~(\ref{eq:wjdef}).
 We set the number of elementary steps between two measurements large enough to avoid autocorrelation effects.  See Table \ref{table:corrtimes} for  details. 
We also performed a binning analysis  as a further test for the independence of the measurements. The statistical error for different bin sizes did not show any change, verifying once more that the sampling distances were large enough. Furthermore, for each system we made several (5 -- 10) runs with randomly chosen starting configurations to independently verify the error bars obtained from the binning analysis.

\begin{table}[htdp]
\begin{center}
\begin{tabular}{r|r|r|r|r|r} 
N  & $\tau_{a.c.}$ & $\Delta n$ &ratio & number of measurements\\
\hline
24 & 22 & 1000 &  45.5 & $10^7$\\
72 & 150 & 1000 & 6.7 &$10^7$\\
96 & 260 & 2000 & 7.7 &$10^7$\\
200 & 970 & 5000 & 5.1& $2\cdot10^6$\\
216 & 1080 & 5000 & 4.6& $2\cdot10^6$\\
384 & 2920 & 20000 & 6.8&$2\cdot10^6$\\
392 & 3340 & 20000 & 6.0&$2\cdot10^6$\\
600 & 7080 & 40000 & 5.6&$10^6$\\
648 & 8100 & 40000 & 4.9&$10^6$
\end{tabular}
\end{center}
\caption{ $\tau_{a.c.}$ autocorrelation times for the two site correlation functions compared to the number of elementary step between two measurements ($\Delta n$).}
\label{table:corrtimes}
\end{table}

We measured diagonal and off-diagonal operators. The spin-spin correlation function, the average of the off-diagonal $P_{k,l}$ can be expressed using the diagonal $ n_{k}^\beta n_{l}^\beta$ operator (where $n_k^\beta$ is the occupation number on site $k$ for the fermion of color $\beta$) as
\begin{equation}
 \langle P_{k,l} \rangle = 20 \langle n_{k}^\beta n_{l}^\beta \rangle - 1 , \label{eq:Pnn}
\end{equation} 
 supposing that the ground state is a singlet wave function -- as it is the case when the hopping Hamiltonian is independent of the colors.
 The measurement of the diagonal $\langle n_k^\beta n_l^\beta \rangle$ correlation functions is quite simple using the importance sampling,  
 \begin{equation}
\left\langle n_k^\beta n_l^\beta \right\rangle_{\text{MC}}  = \frac{N(\{k,l\} \subset \{j^\beta\})}{N_{\text{MC}}} ,
 \end{equation}
where $N(\{k,l\} \subset \{j^\beta\})$ denotes the number of times both $k$ and $l$ sites were occupied with $\beta$ fermion among the $N_{\text{MC}}$ measured configurations. 

With a little more effort one can directly calculate the  off-diagonal $\left\langle \mathcal{P}_{k,l} \right\rangle$  correlation functions as well. Using the fermionic representation for the $\mathcal{P}_{k,l}$ exchange operator, the convenient form that follows the convention of the fermion ordering in the wave function is given as
 \begin{equation}
\mathcal{P}_{k,l} =  \sum_{\alpha\beta} S_\alpha^\beta(k) S_\beta^\alpha(l) 
  = -\sum_{\alpha\beta} f_\alpha^{\dagger}(k) 
  f_\beta^{\dagger}(l) 
  f_\beta^{\phantom{\dagger}}(k) 
  f_\alpha^{\phantom{\dagger}}(l) 
   \label{eq:Pff}
 \end{equation}
In this case one follows the same importance sampling as before, although the measurement itself is more complicated:
\begin{eqnarray}
\left\langle \mathcal{P}_{k,l} \right\rangle 
&=&
\frac{ \sum_{ \{j\},\{\tilde j\}} \bar{w}_{\{j\}} w_{\{\tilde{j}\}} \left\langle \{ j \} \right|   \mathcal{P}_{k,l}   \left| \{ \tilde{j} \}\right\rangle }{  \sum_{ \{j\}} |w_{\{j\}}|^2}  
\nonumber\\ 
& = & 
\frac{
 \sum_{\{j\}} |w_{\{j\}}|^2 \frac{ w_{\{j'\}}}{ w_{\{j\}} } s_{k,l}(\{j\})}
 {
 \sum_{ \{j\}} |w_{\{j\}}|^2
} 
\nonumber\\ 
& = & 
\frac{1}{N_{\text{MC}}} 
 \sum_{\{j\}_{\text{MC}}}  \frac{w_{\{j'\}}}{ w_{\{j\}} } s_{k,l}(\{j\})\;,\label{eq:offdMC}
\end{eqnarray}
where $\{j'\}$ is the configuration that leads to $\{j \}$ by exchanging the color of fermions on sites $k$ and $l$, and the sum in the last equation is over the measured configuration $\{j\}_{\text{MC}}$. Following Eq.~(\ref{eq:Pff}), the sign $s_{k,l}(\{ j\})$ is $1$ if the colors of fermions on sites $k$ and $l$ in the configuration $\{j\}$ are the same, and $-1$ if the colors are different.  We have explicitly verified that the Eq.~(\ref{eq:Pnn}) holds.

Similarly one can calculate  $ \left \langle \mathcal{P}_{ij} \mathcal{P}_{kl} \right\rangle$ as well, here for each $\{j\}$ configuration $|\{ j \}\rangle = \mathcal{P}_{ij} \mathcal{P}_{kl} | \{j' \} \rangle$. For the sign one should take $s_{k,l}(\{ j\})s_{i,j}(\{ j\})$, here we assumed that the sites $i$,$j$,$k$, and $l$ are all different.

\end{appendix}

\bibliographystyle{apsrevtitle}
\bibliography{../refs_su4hc.bib}

\end{document}